\documentstyle[aps,epsf]{revtex}

\begin{document}

\title{Liquid-Gas Phase Transition and Instabilities in 
Asymmetric Two Component Systems}

\author{S.J. Lee}
\address{Department of Physics and Institute of Natural Sciences,
              Kyung Hee University, Suwon, KyungGiDo, Korea}

\author{A.Z. Mekjian}
\address{Department of Physics, Rutgers University, Piscataway NJ 08854}

\maketitle

\begin{abstract}

The liquid-gas phase transition and associated instability in
two component systems are investigated using a mean field theory.
The importance of the role of both the Coulomb force
and symmetry energy terms are studied.
The addition of the Coulomb terms 
bring asymmetry into a mean field and thus
result in important differences
with previous approaches which did not include such terms.
The Coulomb effects modify the chemical instability and
mechanical instability domains
shifting many features away from proton fraction point $y=1/2$
to a value closer to the valley of $\beta$ stability.
These features are discussed in detail.

\end{abstract}



  Thermodynamic properties of two and, more generally, multi-component
systems are of interest in many areas of physics. For example, binary
systems of two components are encountered in nuclei where the two
components are neutrons and protons. 
While light nuclei have $N=Z$, heavy nuclei have $N > Z$
and systems with large neutron excess such as neutron stars
exist in astrophysics.
In condensed matter physics two
component systems appear in liquid $^3$He where the two components are
spin-up and spin-down fluids. Such systems contain a more complex phase
structure than one component systems. This letter is an extension of
previous work concerning the nuclear case which was discussed from both
the fragmentation point of view \cite{ref1,ref2,ref3} 
and from a mean field picture \cite{ref4}.
In the mean field picture we studied the phase surface associated with 
the liquid-gas phase transition incorporating the role of surface and 
Coulomb effects \cite{ref4}. 
Previous mean studies \cite{ref5} of two component systems
were investigations of the role of the symmetry energy on the phase
structure of the liquid-gas phase transition. While the symmetry energy
plays an important role in governing the proton/neutron ratios in the
liquid and gas phases, favoring more symmetric systems in the liquid phase
than the gas phase, the effects of both the Coulomb interaction and
surface energry were also shown to be important. This was first discussed
in Ref.\cite{ref2,ref4} and more recently in Ref.\cite{ref6}. 
Here we will focus on the
stability, metastability and instability of asymmetric systems of protons
and neutrons. Such systems are of interest for radioactive beams studies
which can be used to explore nuclei at the limits of isospin asymmetry.
In a recent article, V. Baran etal \cite{ref7} 
have pointed out some new effects 
associated with two component systems where a new kind of liquid-gas
phase transition associated with chemical instability may appear. This
new type of transition may manifest itself as an isospin distillation 
or fractionization as reported in Ref.\cite{ref8,ref9} 
where the gas phase is enriched
in neutrons relative to the liquid phase. Such observations suggest that
neutron diffusion occurs at the same rate or is more rapid than
fragment production. These observations are based solely on symmetry energy
effects, which make the liquid phase more symmetric,
with the role of the Coulomb interaction not included. 
 The importance of Coulomb effects driven by $Z^2$ terms versus symmetry
energy effects arising from $(N-Z)^2$ terms can be seen by just looking at
the most stable nucleus for a given $A$. Using a liquid drop model, the $Z$ 
with the smallest mass or greatest binding energy is given by
$Z=(A/2)/(1+B_c A^{2/3}/B_s)$ where $B_c$ is the Coulomb coefficient 
$B_c Z^2/A^{1/3}$
and $B_s$ is the symmetry energy coefficient $B_s (N-Z)^2/A$. 
When the Coulomb interaction is turned off $Z$ is simply $A/2$. 
Moreover, the Coulomb interaction was shown \cite{ref2}
to have a very dramatic effect on the specific heat associated with a
first order liquid-gas phase transition, substantially reducing the peak 
previously reported in Ref.\cite{ref3} 
where surface energy effects associated with
one component systems played the dominant role.
Moreover, the Coulomb force introduced another pair of coexistence densities
having larger proton concentration in gas phase than in liquid phase
due to its asymmetry in the mean field \cite{ref4}.

 In symmetric systems, mechanical instability from $dP/d\rho < 0$ and
chemical instability from $d\mu/d\rho < 0$ at fixed $T$ are simply related
since 
\begin{eqnarray}   
            N d\mu = -SdT + VdP        \label{eq1}
\end{eqnarray} 
where $\mu$ is the chemical potential, $S$ the entropy, $P$ the pressure,
$N$ the number of particles and $T$ the temperature. 
Thus $\rho d \mu/d\rho = dP/d\rho$ at fixed $T$ and the boundaries of 
both instabilities are common with $d\mu/d\rho = dP/d\rho = 0$. 
The coexistence curve of the Maxwell pressure $P$
versus density $\rho$ or volume $V$ is determined by the equality of the 
pressure and chemical potentials between the liquid and gas phase at the
same $T$. In the region between the spinoidal curve determined by the
instability line and the coexistence line are metastable regions of 
supercooled liquid and superheated gas where nucleation processes 
can occur. These processes for one component systems were studied in
Ref. \cite{ref10},
where nucleation processes gave a dynamical picture of the 
liquid-gas phase separation. Specifically, classical nucleation theory 
\cite{ref11} looks at the change in the Gibbs free energy with and without
a liquid drop made of $A$ particles. This change is written as
 $\Delta G= (\mu_l-\mu_g) A + 4\pi R^2 \sigma$ where $\mu_l$ 
is the liquid chemical
potential of the drop, $\mu_g$ is the gas chemical potential from 
which the drop is formed, $R$ the radius of the drop and $\sigma$ is the
surface tension. 
A superheated gas has $\mu_g > \mu_l$,
while equilibrium states have $\mu_g = \mu_l$.
This change in Gibbs free energy has a maxima at a
critical size drop whose radius is given by
 $R_c = 2 \sigma/ (\mu_g-\mu_l) \rho_l$, 
where $\rho_l$ is the liquid drop density. 
The $A = \rho_l V$ with $V$ the volume of the drop.
Drops with smaller $R$ evaporate particles thereby lower their free energy,
while drops with bigger $R$ grow by accumulating particles and also
lowering their free energy. Such processes in two component systems of
neutrons and protons with additional Coulomb and symmetry energy effects
have also been partly explored. 
Some preliminary studies of Coulomb effects in
nucleation were done in Ref.\cite{ref12}.
For a two component drop, the $\Delta G$ is
\begin{eqnarray}
 \Delta G = (\mu_{lp} - \mu_{gp}) Z + (\mu_{ln} - \mu_{gn}) N
   + 4\pi R^2 \sigma + \delta E_C + \delta E_{sym}
         \nonumber
\end{eqnarray}
The $\delta E_C$ is the change in Coulomb energy with and without the drop 
and $\delta E_{sym}$ is the change in symmetry energy.

  In two component systems the coexistence line of a one component system
now becomes a surface in $(P,y,T)$, with $y$ the proton fraction. This surface
has several new elements associated with it that are not present in one
component systems. For example, the isolated critical point of a one
component system now becomes a line of critical points with $y=1/2$ the 
critical point of a one component system. Associated with this surface
is also a line of maximal asymmetry in the proton/neutron ratio
and a line of equal concentration in liquid and gas phase \cite{ref4,ref5}. 
The intersection of this $(P,y,T)$ with a plane at fixed $T$ leads 
to loops of $P$ versus $y$ as shown in Fig. 4 of Ref.\cite{ref4}. 
The chemical potentials $\mu_n$ and $\mu_p$
versus $y$ at fixed $T$, $P$ also have interesting behavior as shown in 
Fig. 3 of Ref. \cite{ref4}.
For two component
system, Eq.(\ref{eq1}) is replaced by
\begin{eqnarray}
   y(d\mu_p/d\rho) + (1-y)(d\mu_n/d\rho) = (1/\rho)(dP/d\rho)   \nonumber \\
   y(d\mu_p/dy) +  (1-y)(d\mu_n/dy) = (1/\rho)(dP/dy)       \nonumber
\end{eqnarray}
In the first equation, $y$ and $T$ are held fixed, and in the second equation
$\rho$ and $T$ are held fixed. 
We will use these figures to study the spinoidal instability region and 
its associated surface in $(P,y,T)$.

Some properties of the instability region can best be analyzed  using
some thermodynamic identities. 
We first note that
the pressure and the chemical potentials are usually written 
as functions of $\rho$, $y$ and $T$; $P(\rho,y,T)$ and $\mu_q(\rho,y,T)$.
One important identity arises from the
equation of state $P=P(\rho,y,T)$ which gives
\begin{eqnarray}
 \left(\frac{dy}{d\rho}\right)_{P,T}
  = - \left. \left(\frac{dP}{d\rho}\right)_{y,T} \right/
            \left(\frac{dP}{dy}\right)_{\rho,T}    \label{eq2}
\end{eqnarray}
for the relation of $y$ and $\rho$ at fixed $P$ and $T$.
Since $P$, $T$, $y$ and $\rho$ are connected by an equation of state
$P(\rho,y,T)$,
we can write
\begin{eqnarray}
 \left(\frac{d\mu_q}{dy}\right)_{P,T}
   = \left(\frac{d\mu_q}{dy}\right)_{\rho,T}
    + \left(\frac{d\mu_q}{d\rho}\right)_{y,T}
          \left(\frac{d\rho}{dy}\right)_{P,T}
              \label{eq3} 
\end{eqnarray}
Combining Eqs.(\ref{eq2}) and (\ref{eq3}),
we can write $(dP/d\rho)_{y,T}$ as
\begin{eqnarray}
   \left(\frac{dP}{d\rho}\right)_{y,T}  = 
            \frac{ -(dP/dy)_{\rho,T}(d\mu_q/d\rho)_{y,T} }
            { (d\mu_q/dy)_{P,T} - (d\mu_q/dy)_{\rho,T} }
\end{eqnarray}
Since $(d\mu_q/dy)_{\rho,T}$ is finite, singularites 
in $(d\mu_q/dy)_{P,T}$ (chemical instability)
give rise to zeroes in $(dP/d\rho)_{y,T}$.
The zeroes of $(dP/d\rho)_{y,T}$ form the spinodal surface,
boundary for mechanical instability region.
In two component systems, mechanical and chemical instability regions
are separated for asymmetric cases. 
Using Eq.(\ref{eq2}), 
the chemical instability boundary ($(d\mu_q/dy)_{P,T} = 0$),
determined by setting Eq.(\ref{eq3}) equal to 0, is obtained from
\begin{eqnarray}
 \left(\frac{dP}{d\rho}\right)_{y,T}
    = \left(\frac{dP}{dy}\right)_{\rho,T}
        \frac{(d\mu_q/d\rho)_{y,T}}{(d\mu_q/dy)_{\rho,T}}
     \label{eq4}
\end{eqnarray}
The following features should be noted. 
First, mechanical equilibrium ($(dP/d\rho)_{y,T} = 0$)
is determined by the condition that the left hand side of this equation is
0. Thus the mechanical and chemical instability boundaries are the same 
when $(dP/d y)_{\rho,T} = 0$ since $(d\mu_q/dy)_{\rho,T} \ne 0$ here.
Due to Eq.(\ref{eq2}), $(d\rho/dy)_{P,T} = 0$ too when $(dP/dy)_{\rho,T} = 0$
except on the mechanical instability boundary.
This condition is also related to the equal concentration $y_E$
of liquid and gas phase on the coexistance surface \cite{ref4}.
For systems without Coulomb terms this occurs at $y=1/2$.
Secondly, for chemical equilibrium the right hand side can be obtained
from the equation of state $P=P(\rho,y,T)$ and the behavior of the
proton or neutron chemical potentials $\mu_q=\mu_q(\rho,y,T)$. 
The $q$ will distinguish neutron results from proton results.
We will use Eq.(\ref{eq4}) to explore the boundary of chemical equilibrium.

We begin our study by first considering a simpler system where Coulomb and
surface effects are turned off. Using Eq.(54) of Ref.\cite{ref4} with $C_s=0$
(surface term), $C=0$ (Coulomb term), $x_3=-1/2$ and $\alpha=1$, then
$(dP/d\rho)_{y,T}$ is quadratic in $\rho$ and is given by 
\begin{eqnarray}
 \left(\frac{dP}{d\rho}\right)_{y,T}
  &=& T + \left(\frac{3}{4}\right)t_0\rho + \left(\frac{3}{8}\right)t_3 \rho^2
             + T\left(\frac{1}{4\sqrt{2}}\right)\frac{\lambda^3}{\gamma}\rho
        \nonumber  \\
   & &   - \left[t_0(1/2+x_0)-T\left(\frac{1}{2\sqrt{2}}\right)
              \left(\frac{\lambda^3}{\gamma}\right)\right]
              \left(\frac{\rho}{2}\right)(2y-1)^2
\end{eqnarray}
The last term involving $(2y-1)^2$ does not contribute for symmetric
systems. 
The terms involving $\lambda^3$ arise from degeneracy corrections. 
The mechanical instability points can easily be obtaind by solving
the quadratic equation $(dP/d\rho)_{y,T}=0$ to obtain the two densities that
define the instability region for each $T$ and $y$. These two roots become
a single solution at a specific $T_c(y)$ at each $y$. 
This occurs when $(d^2P/d\rho^2)_{y,T} = 0$.
In one component systems and in symmetric systems with $y=1/2$, 
this is the critical point $T_c$.

\begin{figure}[htb]
\centerline{
\epsfxsize=3in \epsfbox{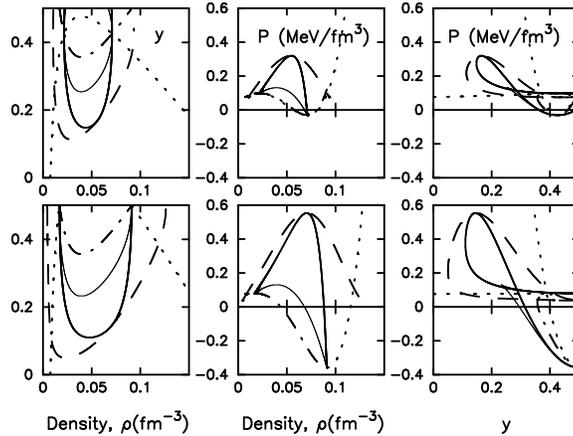}
 }
\caption{
Chemical (thick solid curve) and mechanical (thin solid curve) instability 
curves at temperature $T = 10$ MeV with (upper boxes) 
and without (lower boxes) Coulomb effect of $R = 8$ fm uniform sphere.
The dashed curves are the coexistance curve at $T = 10$ MeV.
The dotted curves are for $(d\mu_p/d\rho)_{y,T} = 0$
and the dash-dotted curves are for $(d\mu_n/d\rho)_{y,T} = 0$.
The equal concentration $y_E = 0.4256$ at $T=10$ MeV.
The curves are the constant $T$ plane cuts of 
the corresponding surfaces in $(\rho,y,T)$ space (left column),
in $(\rho,P,T)$ space (center column), and
in $(P,y,T)$ space (right column).
 }   \label{figtfix}
\end{figure}

\begin{figure}[htp]
\centerline{
\epsfxsize=3in \epsfbox{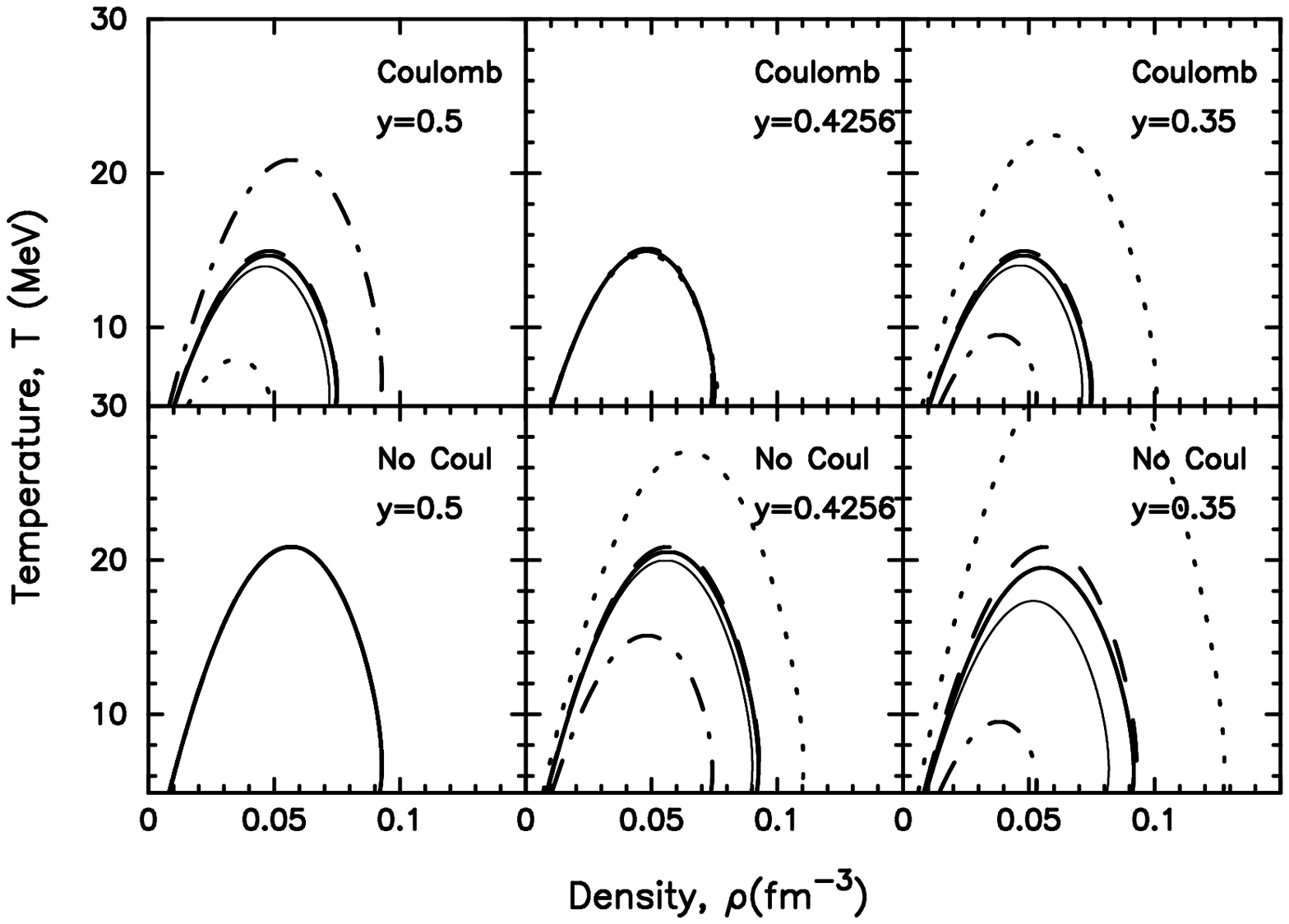}
 }
\caption{
The temperature dependence of the chemical and mechanical instability
curves at fixed proton concentration $y$.
The curves here are the constant $y$ plane cuts of the corresponding 
surfaces in $(\rho,y,T)$ space.
The curves are the same as in Fig. \protect\ref{figtfix}
except the dashed curves. 
With the temperature dependent equal concentration $y_E(T)$,
all the curves coincide as shown by dashed line here
(notice here that $y_E(T) = 1/2$ without Coulomb force).
 }   \label{figyfix}
\end{figure}

The calculations using Eqs.(53) and (54) of Ref. \cite{ref4} 
with degeneracy correction
are shown in Fig. \ref{figtfix} and Fig. \ref{figyfix}.
The mean field force parameters are the same as used in Ref.\cite{ref4}
except for using a larger radius $R = 8$ fm of a uniform sphere 
for a larger Coulomb force.
These figures show that the mechanical instability boundary (thin solid line)
is inside of the chemical instability region (thick solid line) and 
these coincide at the proton concentration of $y_E = 0.4256$ at $T=10$ MeV
(Fig.\ref{figtfix} and upper middle box of Fig.\ref{figyfix}).
The curves of $(d\mu_q/d\rho)_{y,T} = 0$ also coincide
with the instability boundary at $y_E$.
The coexistance boundary coincides with 
the chemical instability boundary
at the line of critical points 
(peaks in Fig.\ref{figtfix} for a $T=10$ MeV plane cut)
but not with the mechanical instability boundary .
We can see from Fig.\ref{figyfix} that the chemical and mechanical 
instability boundaries coincide at the proton concentration $y_E(T)$
which is on the line of equal concentration on the coexistance 
surface \cite{ref4}.
While all the curves coincide at for $y=1/2$ (without Coulomb)
as shown 
in the lower left box of Fig.\ref{figyfix}, 
the top middle box of Fig. \ref{figyfix} with Coulomb exhibit
a separation between these curves due to the temperature dependence
of $y_E(T)$. 
All curves in the upper middle box of Fig. \ref{figyfix}
coincide at $T= 10$ MeV since $y_E = 0.4256$ at this temperature.
From Fig.\ref{figyfix} and the left column of Fig.\ref{figtfix},
we can see that all the four surfaces in $(\rho,y,T)$ space coincide
at the line of temperature dependent $y_E(T)$
and that the mechanical instability surface is inside of 
the chemical instability surface.
Upper Fig. \ref{figyfix} and upper left box of Fig.\ref{figtfix}
also shows that the role of $(d\mu_p/d\rho)_{y,T}=0$
and $(d\mu_n/d\rho)_{y,T}=0$ are exchanged
when $y$ crosses $y_E$.

To keep equations simple in the dicussions of various new features
in two component system with Coulomb force,
we will drop the degeneracy corrections.
However, they are included in the 
actual calculations done and in the fiqures (Figs. \ref{figtfix}
and \ref{figyfix}) presented here.
In this limit $P=\rho T-a_0(y)\rho^2+2a_3\rho^3$   
where we have followed the notation 
of Ref. \cite{ref10} with $P$ now generalized to the case of 
asymmetric systems. 
The generalization changes $a_0$ to
 $a_0(y)=a_0 [1-(2/3) (1/2 + x_0) (2y-1)^2]$ 
with $a_0=-(3/8)t_0$ and $a_3=t_3/16$. 
Then $dP/d\rho=0$ at     
\begin{eqnarray}
     \rho_{\pm}=\left(a_0(y) \pm \sqrt{a_0^2(y)-6a_3 T}\right)/(6a_3)
\end{eqnarray}
The two roots are equal at $T_c(y) = a_0^2(y)/(6a_3)$ and are
$\rho_c(y) = \rho_+ = \rho_- = a_0(y)/(6a_3)$. At this point the pressure is 
$P_c(y) = a_0^3(y)/(108 a_3^2)$. Under the same limits the chemical
potentials are given by 
\begin{eqnarray}
     \mu_q = T\ln[(\lambda^3/\gamma)(\rho/2 \pm (2y-1)\rho/2)]
           - 2a_0 \rho \pm (4/3)(1/2 + x_0) a_0 (2y-1)\rho + 3a_3\rho^2 
\end{eqnarray}
with the upper sign for protons and the lower sign for neutrons
and $\gamma=2$ for the spin up and down degeneracy. 
This equation for $\mu_q = \mu_q(\rho,y,T)$ when combined 
with the equation of state $P(\rho,y,T)$ can be used to calculate 
the chemical potentials as a function $\mu_q = \mu_q(P,y,T)$. 
Boundaries of the chemical instability region can
also be determined using the condition $d\mu_q/dy = 0$ at constant $T$, $P$. 
Substituting the simplified equation of state and expression for the
chemical potential into Eq.(\ref{eq4}) leads to a cubic equation
 $c_3\rho^3+c_2\rho^2+c_1\rho+c_0=0$ 
with $c_3=32 (1/2+x_0) a_0 a_3 (1-y)y$,
 $c_2=[6a_3 T-(32/3) (1/2+x_0) a_0^2 (1-y)y]$,
 $c_1=- (4/3) (1-x_0) a_0 T$, $c_0=T^2$.
Chemical instability occurs when three real
roots exist, two of which will be postive, one negative. 
One negative root is present because $c_0/c_3$ is positive.
At $y=1/2$, the two positive solutions are the same as that for mechanical 
instability.
When $y$ is no longer 1/2, the mechanical instability
and chemical instability regions are no longer the same. 
This result is shown in the lower part of Fig.\ref{figyfix} 
for various choices of $y$ 
with the degenercy correction
of Ref. \cite{ref4}. 
It should also be noted that the coefficients of the cubic equation 
have a certain symmetry arising 
from the form in $y$ which is $y(1-y)$, i.e., interchanging
$y$ (protons) and $1-y$ (neutrons) leaves the solution invariant
(see curves for $(d\mu_q/d\rho)_{y,T} = 0$ in Figs. \ref{figtfix} and 
\ref{figyfix}). 
This result is no longer true when Coulomb interactions are included. 
We now turn our attention to the simple EOS and $\mu_q$, 
but now with Coulomb effects included
in a simplified form which will illustrate its qualitative effects.

When Coulomb effects are included in the approximation of 
a uniformly charged sphere of fixed radius $R$, 
the pressure equation contains 
an additional term $Cy^2\rho^2$
and the proton chemical potential contains an extra term $2Cy\rho$.
The $C$, which depends on R, is taken to be a constant.
For this choice, we can consider R as an effective range of the Coulomb
force which is now independent of the density.
This simplification reduces the complexity of many results.
To make the Coulomb effects more visible in our figures 
(Figs. \ref{figtfix} and \ref{figyfix}),
we use a large value of $C = 231$ MeV/fm$^3$ 
which is appropriate for systems near U.  
These additions to $P$ and $\mu_p$ change the equation for
chemical and mechanical instability.
Specifically the coefficients of the cubic equation for
chemical instability for both protons and neutrons are now
 $c_3 = [32 (1/2+x_0) a_0 + 12 C] a_3 y(1-y)$,
 $c_2 = 6a_3 T - [(32/3) (1/2+x_0) a_0^2 + (8/3) (1-x_0) a_0 C] y(1-y)$,
 $c_1 = [-(4/3) (1-x_0) a_0 + 2Cy] T$, $c_0 = T^2$.

Some qualitative features of the effects of such a Coulomb behavior
are the following (see Fig. \ref{figtfix} and Fig.\ref{figyfix} also).
1) The mechanical and chemical instability curves are the same
when $(dP/d y)_{\rho,T} = 0$ which occurs at
 $y_E = 0.5/(1 + C/[(8/3)(1/2+x_0)a_0])$ 
and thus no longer at $y_E=0.5$ when $C \ne 0$.
This equation for $y$ has a similar structure to the equation
that gives the $Z$ which has the maximum binding energy
at fixed $A$ (valley of $\beta$ stability) obtained
from the liquid drop mass formulae or from Eq.(55) of Ref.\cite{ref4}.
With higher order $T$ dependence coming from degeneracy corrections,
 $y_E$ depends on $T$ and $y_E = 0.4256$ for $x_0=1/2$ at $T = 10$ MeV.
Explicitly, with first order degeneracy corrections,
 $y_E = 0.5/(1 + C/[(8/3) (1/2+x_0) a_0 + T \lambda^3/(2\sqrt{2} \gamma)])$.
2) For both larger and smaller values of $y$ from $y_E$,
the chemical instability curve is outside the mechanical
instability curve as shown in Fig.\ref{figtfix} and \ref{figyfix}.
The mechanical instability has its peak in $T$ versus $\rho$ 
at $T_c(y) = [a_0 - (2/3) (1/2+x_0) a_0 (2y-1)^2 - C y^2]^2/(6 a_3)$ which
is maximum at $y = y_E$ and coincide with chemical instability at $y_E$.
3) The maximum critical $T$ obtained from the peak in the chemical 
equilibrium surface of $T$ versus $\rho$, occurs at $y \ne 0.5$,
but is now shifted to $y=y_E(T)$.
4). At $y=y_E(T)$, the equilibrium curve or binodal curve of $P$
versus $y$ obtained by cutting the binodal surface at a given $T$
meets at a point, i.e., the liquid and gas phase have the
same proton concentration (see Fig.4 of Ref.\cite{ref4}).
For $y > y_E$ the gas phase is proton richer than the liquid phase 
when compared to a situation without Coulomb effects \cite{ref4}.
When Coulomb effects are turned off the binodal section meets
at $y=1/2$.
At $y = y_E \ne 1/2$ with Coulomb effects included
a two component system will behave as a one component system,
keeping the proton/neutron ratio the same in both
liquid and gas phases.

In summary,
we have presented calculations based on a mean field theory
with asymmetry which include the role of Coulomb force 
on the liquid-gas phase
instability regions for both the mechanical (spinodal) surface
and the chemical (diffusive) surface.
The Coulomb force modifies these surfaces so that their peaks
are no longer at proton fraction $y=1/2$
but shifted to a smaller value called $y_E$ 
which is close to the proton fraction in the valley 
of $\beta$-stability.
The mechanical instability surface is always inside the
chemical instability surface and the two surfaces are 
the same at the equal concentration $y_E$ 
(allowing for $T$ dependence in $y_E$).
At $y_E$, the peak in the chemical instability surface and
binodal equilibrium surface also coincide.
For systems with $y \ne y_E$, the peak in the $T$ versus $\rho$
curve for chemical instability and chemical equilibrium --
binodal or coexistance curve -- coincide,
but the peak for mechanical or spinodal instability is
below that of the other two.
At $y = y_E$, a phase transition in a two component system
behaves the same as a one-component system,
keeping the proton fraction at $y_E$ in both phases
during the entire transformation.

This work was supported 
in part by the DOE Grant No. DE-FG02-96ER-40987
and in part by Grant No. R05-2001-000-00097-0 from the Basic Research Program 
of the Korea Science and Engineering Foundation.

\end{document}